\documentclass[twocolumn,aps,prl,superscriptaddress,tightenlines]{revtex4-2}
\usepackage{graphicx}
\usepackage{dcolumn}
\usepackage{bm}
\usepackage{epsfig}
\usepackage{amsmath}
\usepackage{amsfonts}
\usepackage{amssymb}
\usepackage{empheq}
\usepackage{color}
\usepackage{bbm}
\usepackage[caption=false]{subfig}
\usepackage{placeins}
\usepackage{soul}
\usepackage{physics}
\usepackage{mathrsfs}
\usepackage{mathtools}
\usepackage{changes} 
\definechangesauthor[name={Per cusse}, color=red]{per} 
\usepackage[colorlinks,citecolor=blue]{hyperref}
\usepackage{todonotes}
\usepackage{orcidlink}
\usepackage{centernot}

\newcommand{\newg}{\raisebox{0.28ex}{\ensuremath{\mathrm{g}}}}

\makeatletter
\newcommand*\bigcdot{\mathpalette\bigcdot@{.8}}
\newcommand*\bigcdot@[2]{\mathbin{\vcenter{\hbox{\scalebox{#2}{$\m@th#1\bullet$}}}}}
\makeatother

\begin{document}
	\title{Towards High-Brightness Perfect Photon Blockade}
	\author{Zhi-Guang Lu\,\orcidlink{0009-0007-4729-691X}}
	\affiliation{School of Physics and Institute for Quantum Science and Engineering, Huazhong University of Science and Technology and Wuhan Institute of Quantum Technology, Wuhan 430074, China}
	
	\author{Xin-You L\"{u}}\email{xinyoulu@hust.edu.cn}
	\affiliation{School of Physics and Institute for Quantum Science and Engineering, Huazhong University of Science and Technology and Wuhan Institute of Quantum Technology, Wuhan 430074, China}
	
	\date{\today}
	\begin{abstract}
           Single-photon sources with high single-photon purity and high brightness are key elements of many future quantum technologies. While photon blockade (PB) is widely exploited in the development of such sources, achieving the coexistence of high purity and high brightness remains a long-standing challenge. Here, we identify a novel mechanism for high-brightness PB and demonstrate that near-ideal purity and near-ideal brightness can be simultaneously achieved in an extended nondegenerate two-photon Jaynes-Cummings model with two-body and three-body interactions. This mechanism is underpinned by a distinctive energy-level structure arising from the combined action of the two interactions. The energy levels in the multi-excitation manifold essentially retain a harmonic ladder of degenerate doublets, whereas in the single-excitation subspace the doublet degeneracy is lifted, with a finite splitting between the two levels. Consequently, when one bosonic mode is driven by a coherent continuous-wave pump, the former degeneracy enables the other bosonic mode to exhibit near-perfect PB even in the strong driving regime, while the latter splitting allows the mean photon number of that mode to approach unity. Our proposed scheme overcomes the outstanding challenge and offers a promising pathway toward realizing ideal single-photon sources. 
	\end{abstract}
	\maketitle
	
	\emph{Introduction}.---Single-photon purity and brightness are key performance metrics of a high-quality single-photon source \cite{shields2007semiconductor, santori2002indistinguishable, Oxborrow01052005, RevModPhys.87.347, PhysRevLett.81.1110}, which constitutes a fundamental building block for quantum communication \cite{RevModPhys.84.777}, quantum computing \cite{RevModPhys.79.135}, and quantum cryptography \cite{RevModPhys.81.1301, PhysRevLett.85.1330}. A common way to quantify single-photon purity and brightness is via the equal-time second-order correlation $g^{(2)}(0)=\langle:\!\hat{n}^2\!\!:\rangle/\langle\hat{n}\rangle^2$ and the mean photon number $\langle\hat{n}\rangle$, respectively, where $\hat{n}$ is the photon-number operator and colons indicate normal ordering. For an ideal single-photon state, one has $g^{(2)}(0)=0$ and $\langle\hat{n}\rangle=1$, corresponding to ideal single-photon purity and unit brightness. Notably, the photon blockade (PB) \cite{PhysRevLett.79.1467}, the suppression of subsequent photons following the absorption of the first, provides a very promising mechanism for integrable and scalable single-photon sources and has attracted considerable theoretical and experimental attention \cite{birnbaum2005photon, science1152261, PhysRevLett.121.043601, PhysRevLett.121.043602, PhysRevLett.106.243601, PhysRevLett.118.133604, PhysRevLett.107.053602, PhysRevLett.114.233601, PhysRevLett.103.150503, PhysRevLett.125.197402, PhysRevA.65.063804, PhysRevLett.107.063601, PhysRevLett.109.193602, PhysRevLett.123.013602, PhysRevLett.130.243601, PhysRevLett.127.240402, PhysRevLett.134.013602}. 
    
    While the PB has been realized in a variety of physical systems, its implementations involve diverse mechanisms and schemes in different parameter regimes. Specifically, the PB is commonly classified into the conventional \cite{PhysRevLett.79.1467} and unconventional \cite{PhysRevA.83.021802, PhysRevLett.104.183601, PhysRevA.100.063817, PhysRevA.96.053810} mechanisms. The former relies on an anharmonic energy spectrum induced by strong nonlinearity, whereas the latter arises from an interference effect between excitation pathways, which can occur even in weakly nonlinear systems. In principle, both mechanisms can achieve near-complete suppression of the equal-time second-order correlation, i.e., $g^{(2)}(0)\approx0$, also known as a near-perfect PB \cite{PhysRevLett.121.043602}. It is worth noting that most previous studies have focused on a weak driving regime, which has been commonly considered in the two mechanisms. This mainly is because strong driving populates higher multiphoton manifold, which not only complicates analytical treatments of photon correlations, but also tends to greatly worsen a PB effect \cite{PhysRevA.100.063817}. By contrast, in the weak driving limit, the Hilbert space can be truncated to a low-photon manifold and quantum jumps become extremely rare \cite{PhysRevA.96.053810}, thereby simplifying analytical treatments and favoring the PB. However, this trade-off manifests as a vanishing brightness, i.e., $\langle\hat{n}\rangle\approx0$. Hence, achieving the coexistence of high purity and high brightness remains an outstanding challenge to date, prompting the fundamental question: {\it Is it possible to generate a steady-state single-photon source that simultaneously exhibits both properties?}

    In this Letter, we identify a novel mechanism for high-brightness PB and demonstrate that the coexistence of high purity and high brightness can be achieved in an extended nondegenerate two-photon Jaynes-Cummings (JC) model \cite{PhysRevA.40.5116, PhysRevA.45.8121, PhysRevA.50.5116}, in which a two-level system (TLS) interacts with two cavity modes via a two-photon JC process, while the cavity modes are coupled through photon tunneling. In the regime where the two-photon process dominates, the energy levels of this system in the multi-excitation manifold essentially feature a harmonic ladder of degenerate doublets, whereas in the single-excitation subspace the doublet degeneracy is lifted by the photon tunneling, with a finite splitting between the two levels. As a consequence, when one cavity mode is driven by a coherent cw pump, the former degeneracy effectively suppresses the multiphoton populations of the other cavity mode even in the strong driving regime, while the latter splitting enables a coherent population transfer between single-photon states of the two cavity modes. This shows that such coexistence is indeed achievable within our proposed scheme. Moreover, we derive an effective reduced master equation that not only provides a mathematically convenient description, but also naturally elucidates the underlying physical mechanism.

    \begin{figure}[t]
        \includegraphics[width=8.5cm]{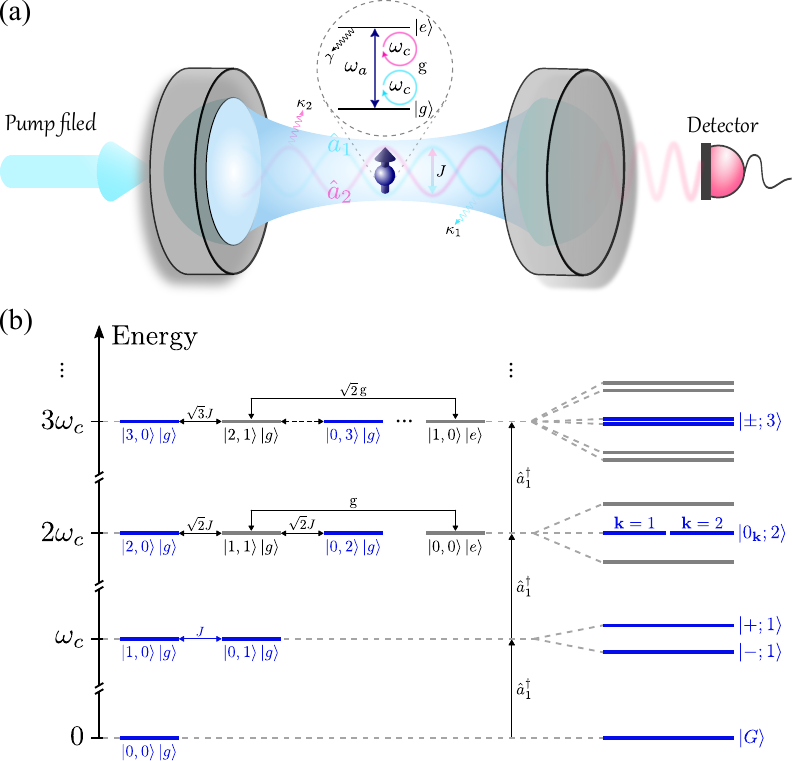}
        \caption{(a) Setup of the extended nondegenerate two-photon Jaynes-Cummings model. (b) Energy spectrum of the system Hamiltonian at $\omega_a=2\omega_c$ in the subspaces containing zero, one, two, and three excitations, shown in the uncoupled basis (left) and the diagonal basis (right). The up arrows illustrate photon transition processes of the driven cavity mode induced by the pump field (cyan arrow), and blue lines highlight the core energy levels of the system.
        }\label{fig1}
    \end{figure}

	\emph{Model and Hamiltonian}.---We consider the extended nondegenerate two-photon JC model, in which a TLS is coupled to two linearly interacting cavity modes, as schematically illustrated in Fig.\,\ref{fig1}(a). In the absence of external driving and dissipation, the Hamiltonian of the system can be expressed as $\hat{H}_{\rm sys}=\hat{H}_0+\hat{H}_{\rm int}$. Here, the free Hamiltonian reads $\hat{H}_0=\omega_a\hat{\sigma}_{}^{\dagger}\hat{\sigma}_{}^{}+\omega_c(\hat{a}_1^\dagger \hat{a}_1^{}+\hat{a}_2^\dagger \hat{a}_2^{})$, where $\hat{\sigma}=\ketbra{g}{e}$ is the lowering operator for the TLS with transition frequency $\omega_a$ and $\hat{a}_\mathbf{k}$ is the annihilation operator for the $\mathbf{k}$-th cavity mode with resonant frequency $\omega_c$; the interaction Hamiltonian takes the form
    \begin{align}
    \hat{H}_{\rm int}=J(\hat{a}^\dagger_1\hat{a}_2^{}+\hat{a}_2^\dagger\hat{a}_1^{})+\newg (\hat{a}_1^{}\hat{a}_2^{}\hat{\sigma}_{}^{\dagger}+\hat{\sigma}_{}^{}\hat{a}_1^\dagger\hat{a}_2^\dagger),\label{Eq1}
    \end{align}
    where the first term is a two-body interaction (2BI) with hopping amplitude $J$, corresponding to photon tunneling between the two cavity modes, and the second term represents a three-body interaction (3BI) with coupling strength $\newg$, describing two-photon JC process involving the TLS and the two cavity modes.

    The system evolves in a discrete, infinite-dimensional Hilbert space in which the uncoupled $(\newg=J=0)$ eigenvectors are labeled as $|n\rangle_1\otimes|m\rangle_2\otimes\ket{g\,(e)}\equiv\ket{n,m}\ket{g\,(e)}$, where $\ket{g\,(e)}$ is the ground (excited) state of the TLS and $|n\rangle_\mathbf{k}$ represents the Fock state with $n$ photons on the $\mathbf{k}$-th cavity mode. A crucial feature of the system Hamiltonian is the conservation of the weighted excitation number $\hat{N}=2\hat{\sigma}^\dagger\hat{\sigma}^{}+\hat{a}_1^\dagger\hat{a}_1^{}+\hat{a}_2^\dagger\hat{a}_2^{}$, and thus the system can be diagonalized in each subspace $\mathcal{H}_n$ containing exactly $n$ excitations. Notice that the spectrum of $\hat{H}_{\rm int}$ is symmetric around zero due to the chiral symmetry \cite{supp}. In the resonant case (i.e., $\omega_a=2\omega_c$), the interaction Hamiltonian plays a decisive role in determining the energy spectrum and dynamics of the system, as a consequence of $\hat{H}_0/\omega_c=\hat{N}$. For $\newg\neq0$ and $J=0$, the interaction (3BI) Hamiltonian possesses a pair of degenerate zero-energy eigenstates, $\ket{n,0}\ket{g}$ and $\ket{0,n}\ket{g}$, in the subspace $\mathcal{H}_{n}$ for each $n\ge1$. As a result, the energy spectrum of the system Hamiltonian features a harmonic ladder of the degenerate doublets. 
    However, once the 2BI is switched on (i.e., $J\neq 0$), the zero-energy degeneracy in each odd-excitation subspace is lifted, thereby splitting into a pair of eigenstates $\ket{\pm;2s+1}\in\mathcal{H}_{2s+1}$ with nonzero energies $\pm \Delta E_s$, i.e., $\hat{H}_{\rm int}\ket{\pm;2s+1}=\pm\Delta E_s\ket{\pm;2s+1}$. By contrast, in each even-excitation subspace the zero-energy degeneracy survives and is realized as a pair of new degenerate eigenstates $\ket{0_{\mathbf{k}};2s}\in\mathcal{H}_{2s}$ with $\mathbf{k}\in\{1,2\}$, which satisfy $\hat{H}_{\rm int}\ket{0_\mathbf{k};2s}=0$, as also illustrated in Fig.\,\ref{fig1}(b). For concreteness, one finds $\ket{\pm;1}\propto(\ket{1,0}\pm\ket{0,1})\ket{g}$ and $\ket{0_\mathbf{k};2}\propto\ket{2\delta_{\mathbf{k},1},2\delta_{\mathbf{k},2}}\ket{g}-\sqrt{2}(J/\newg)\ket{0,0}\ket{e}$, where $\delta_{\mathbf{k},\mathbf{l}}$ is the Kronecker symbol. Notably, in the limit of $J/\newg\ll1$, the above energy splitting in the subspace $\mathcal{H}_{2s+1}$ can be obtained analytically as \cite{supp}
    \begin{align}\label{Eq2}
        \Delta E_0/J=1,\ \ \Delta E_{s}/J\simeq \mathscr{T}_{2s}(J/\newg)^{2s}\simeq0\quad(s\ge1),
    \end{align}
    where $\mathscr{T}_{2s}=(2s+1)!!/(2s)!!$. This clearly shows that, apart from a finite splitting induced by the 2BI in the single-excitation subspace, the energy spectrum in multi-excitation manifold $\mathcal{H}_2\oplus\mathcal{H}_3\oplus\cdots\oplus\mathcal{H}_\infty$ essentially retains the harmonic ladder of degenerate doublets, which constitutes a distinctive energy-level structure of our system. Throughout this Letter, we focus on the resonant case so as to maximize the effect of the 3BI. 

    To achieve ideal steady-state single-photon sources, we assume cavity mode $\hat{a}_1$ is driven by a coherent cw pump with frequency $\omega_{p}$ and amplitude $\Omega$. In a rotating frame defined by the unitary operator $\text{exp}(-i\omega_pt\hat{N})$, the total Hamiltonian of the system then reads
    \begin{align}\label{Eq3}
        \hat{H}_{\rm tot} = -\Delta\hat{N}+\hat{H}_{\rm int}+\hat{H}_{\rm d},\quad \hat{H}_{\rm d}=\Omega(\hat{a}_1^{}+\hat{a}_1^\dagger),
    \end{align}
    where $\Delta=\omega_p-\omega_c$ represents the pump-cavity detuning. In the presence of dissipation, the evolution of the density matrix $\hat{\rho}$ is given by the full master equation (FME)
    \begin{align}\label{Eq4}
        \dot{\hat{\rho}}=-i[\hat{H}_{\rm tot},\hat{\rho}]+\kappa_1\mathcal{D}[\hat{a}_1]\hat{\rho}+\kappa_2\mathcal{D}[\hat{a}_2]\hat{\rho}+\gamma\mathcal{D}[\hat{\sigma}]\hat{\rho},
    \end{align}
    where $\mathcal{D}[\hat{o}]\hat{\rho}=\hat{o}\hat{\rho}\hat{o}^\dagger-(\hat{o}^\dagger\hat{o}\hat{\rho}+\hat{\rho}\hat{o}^\dagger\hat{o})/2$ is the Lindblad superoperator. This describes the dissipation of the cavity mode $a_1\,(a_2)$ and TLS with the decay rates $\kappa_1\, (\kappa_2)$ and $\gamma$, respectively. By solving the FME (\ref{Eq4}), one can obtain the mean photon number $\expval{\hat{n}_\mathbf{k}}$ and equal-time second-order correlations $g^{(2)}_\mathbf{k}(0)=\langle\hat{n}_\mathbf{k}(\hat{n}_\mathbf{k}-1)\rangle/\langle\hat{n}_\mathbf{k}\rangle^2$ in the steady state $\hat{\rho}_{\rm ss}$, where $\langle{\bigcdot}\rangle\coloneqq\mathrm{Tr}[\mkern2mu\bigcdot\mkern2mu\hat{\rho}_{\rm ss}]$ and $\hat{n}_\mathbf{k}=\hat{a}^\dagger_\mathbf{k}\hat{a}^{}_\mathbf{k}$.
    
\begin{figure}
        \includegraphics[width=8.5cm]{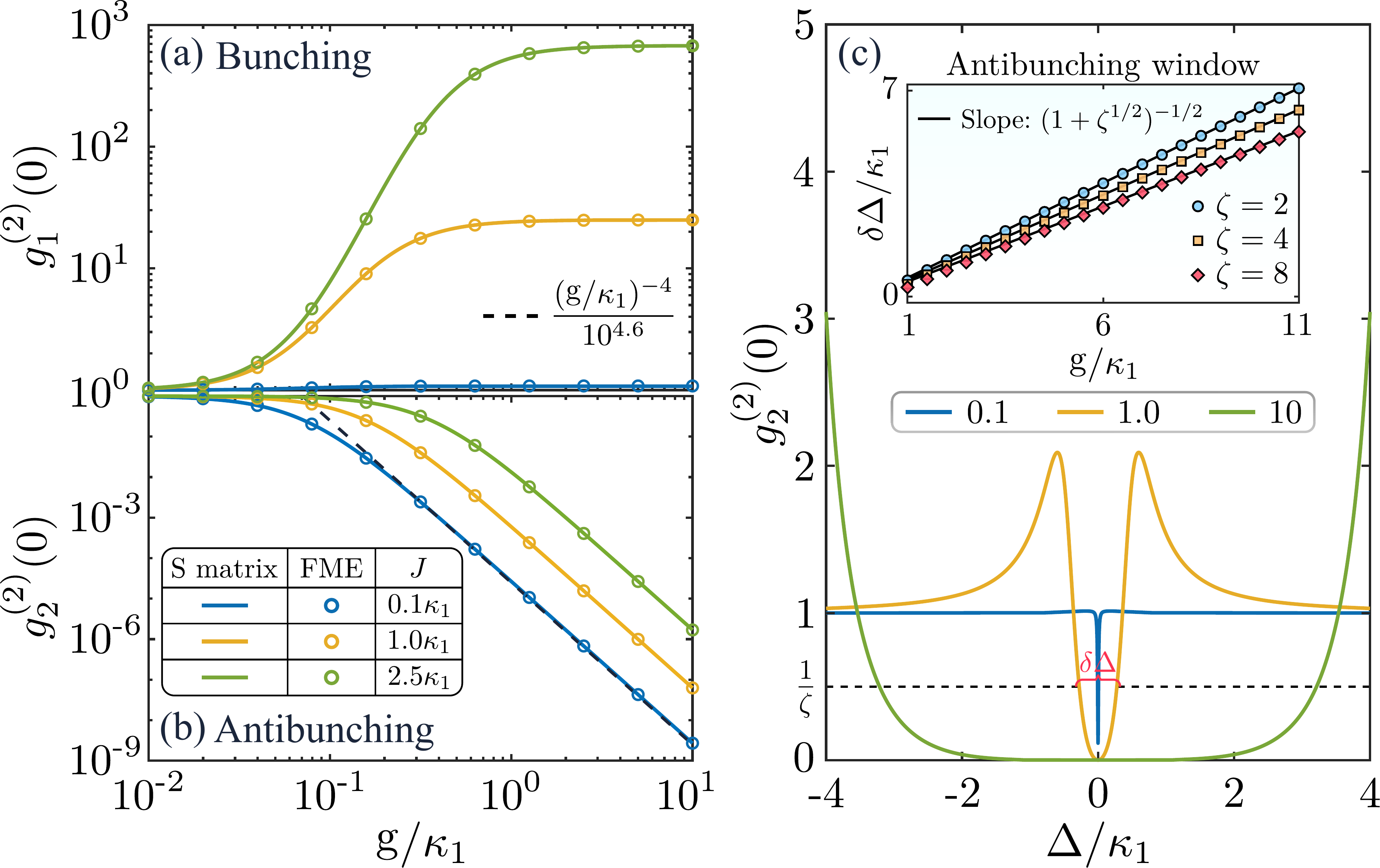}\\
        \caption{Equal-time second-order correlations $g^{(2)}_{\bf k}(0)$ of (a) the driven and (b) the undriven cavity modes versus the coupling strength $\newg/\kappa_1$ for $J=\{0.1,1,2.5\}\kappa_1$ and $\Delta=0$. The dashed black line shows the biquadratic scaling law. (c) Dependence of $g_{2}^{(2)}(0)$ on the pump detuning $\Delta/\kappa_1$ for $\newg=\{0.1,1,10\}\kappa_1$ and $J=0.1\kappa_1$, with the inset showing the variations of $\delta\Delta$ versus the coupling strength $\newg/\kappa_1$ for $\zeta=2,4,8$. Notice that the discrete data points in panels (a)--(b) are numerical solutions to the FME in the weak driving amplitude $\Omega=10^{-4}\kappa_1$, and the solid curves are obtained from the analytic expression (\ref{Eq5}). The remaining parameters are $\kappa_2=\kappa_1$ and $\gamma=0.01\kappa_1$.
        }\label{fig2}
    \end{figure}

    \emph{Analytic solutions in the weak driving limit}.---We first employ the scattering matrix (S matrix) method \cite{PhysRevA.82.063821, PhysRevA.91.043845, PhysRevA.92.053834, PhysRevLett.122.243602, PhysRevA.108.053703} to analytically derive higher-order correlations in the weak driving limit, which agree perfectly with the numerical results based on the FME. In the limit of $\Omega/\kappa_1\to0$, $g^{(2)}_\mathbf{k}(0)$ takes the analytic form \cite{supp}
    \begin{align}\label{Eq5}
        g_{\mathbf{k}}^{(2)}(0)=\abs{1+\frac{\delta_{\mathbf{k},1}\mathcal{C}_2-\delta_{\mathbf{k},2}}{(1+\mathcal{C}_2+\mathcal{C}_3)/\mathcal{C}_3}}^2,
    \end{align}
    where $\mathcal{C}_2=4J^2[(\kappa_1-2i\Delta)(\kappa_2-2i\Delta)]^{-1}$ is the two-body parameter and $\mathcal{C}_3=4\mkern0.5mu\newg^2[(\kappa_1+\kappa_2-4i\Delta)(\gamma-4i\Delta)]^{-1}$ is the three-body parameter. At zero detuning $\Delta=0$, since $\mathcal{C}_{2},\mathcal{C}_3\in\mathbb{R}^+$, one always has $g_1^{(2)}(0)>1$ and $g_2^{(2)}(0)<1$. This demonstrates that the driven and undriven cavity modes invariably exhibit photon bunching [see Fig.\,\ref{fig2}(a)] and antibunching [see Fig.\,\ref{fig2}(b)], respectively. Physically, when $\Delta=0$, there is a two-photon resonance with the transitions $\ket{0,0}\ket{g}\equiv\ket{G}\to|0_{\bf k};2\rangle$, whereas the single-photon transitions $\ket{\rm G}\to\ket{\pm;1}$ are detuned by $\pm \Delta E_0$. Then, as the 3BI strength increases, the two degenerate eigenstates $\ket{0_{\bf k};2}$ become increasingly distinct in their state composition, with $\braket{0_{\bf 1};2}{0_{\bf 2};2}\simeq2(J/\newg)^2$, thereby progressively suppressing the transition $\ket{0_{\bf 1};2}\to\ket{0_{\bf 2};2}$. Especially in the limit of $J/\newg\ll1$, it follows directly that $\ket{0_{\bf k};2}\simeq\ket{2\delta_{\mathbf{k},1},2\delta_{\mathbf{k},2}}\ket{g}$. This thus indicates that, in the driven cavity mode, absorption of the first photon facilitates that of the second, resulting in photon bunching. By contrast, in the undriven cavity mode, once the first photon is absorbed, absorption of the second is strongly suppressed because the transition $\ket{0_{\bf 1};2}\to\ket{0_{\bf 2};2}$ is effectively forbidden, leading to near-perfect PB. Notably, $g_2^{(2)}(0)$ exhibits biquadratic scaling with the 3BI strength [see dashed line of Fig.\,\ref{fig2}(b)], approximately given by
    \begin{align}\label{Eq6}
        g_2^{(2)}(0)\simeq [(1+\mathcal{C}_2)/\mathcal{C}_3]^2\propto(1/\newg)^{4},
    \end{align}
    in stark contrast to the quadratic scaling arising from the anharmonicity of a single-mode JC model or a nonlinear cavity \cite{PhysRevA.110.063705}. It is worth noting that a larger energy splitting $\Delta E_0$ further suppresses the absorption of the first photon, which leads to an overall enhancement of $g_\mathbf{k}^{(2)}(0)$ as the 2BI increases [see Figs.\,\ref{fig2}(a) and \ref{fig2}(b)].

    We next study the impact of detuning on the photon antibunching in the undriven cavity mode. As seen from Fig.\,\ref{fig2}(c), the optimal antibunching occurs at zero detuning, whereas the antibunching gradually deteriorates as the detuning $\abs{\Delta}$ increases. To properly quantify this impact, we introduce an antibunching detuning window $\delta\Delta$, defined as the interval $\Delta\in[-\delta\Delta/2, \delta\Delta/2]$ within which $g_2^{(2)}(0)<1/\zeta<1$, where $\zeta$ denotes a threshold parameter characterizing the strength of antibunching. Notably, in the regime where $J/\newg\ll1$ and $\Delta/\newg=\order{1}$, we prove that the antibunching window exhibits a linear dependence on the 3BI strength, i.e., $\delta\Delta\simeq\newg(1+\sqrt{\zeta})^{-1/2}$ \cite{supp}, as shown in the inset of Fig.\,\ref{fig2}(c). This further shows that a stronger 3BI not only induces near-perfect PB, but also enhances the robustness of PB (i.e., strong antibunching) against frequency disorder. Accordingly, we focus on the zero detuning ($\Delta=0$) in the following discussion, as this optimally favors photon blockade in our system.

\begin{figure}
        \includegraphics[width=8.5cm]{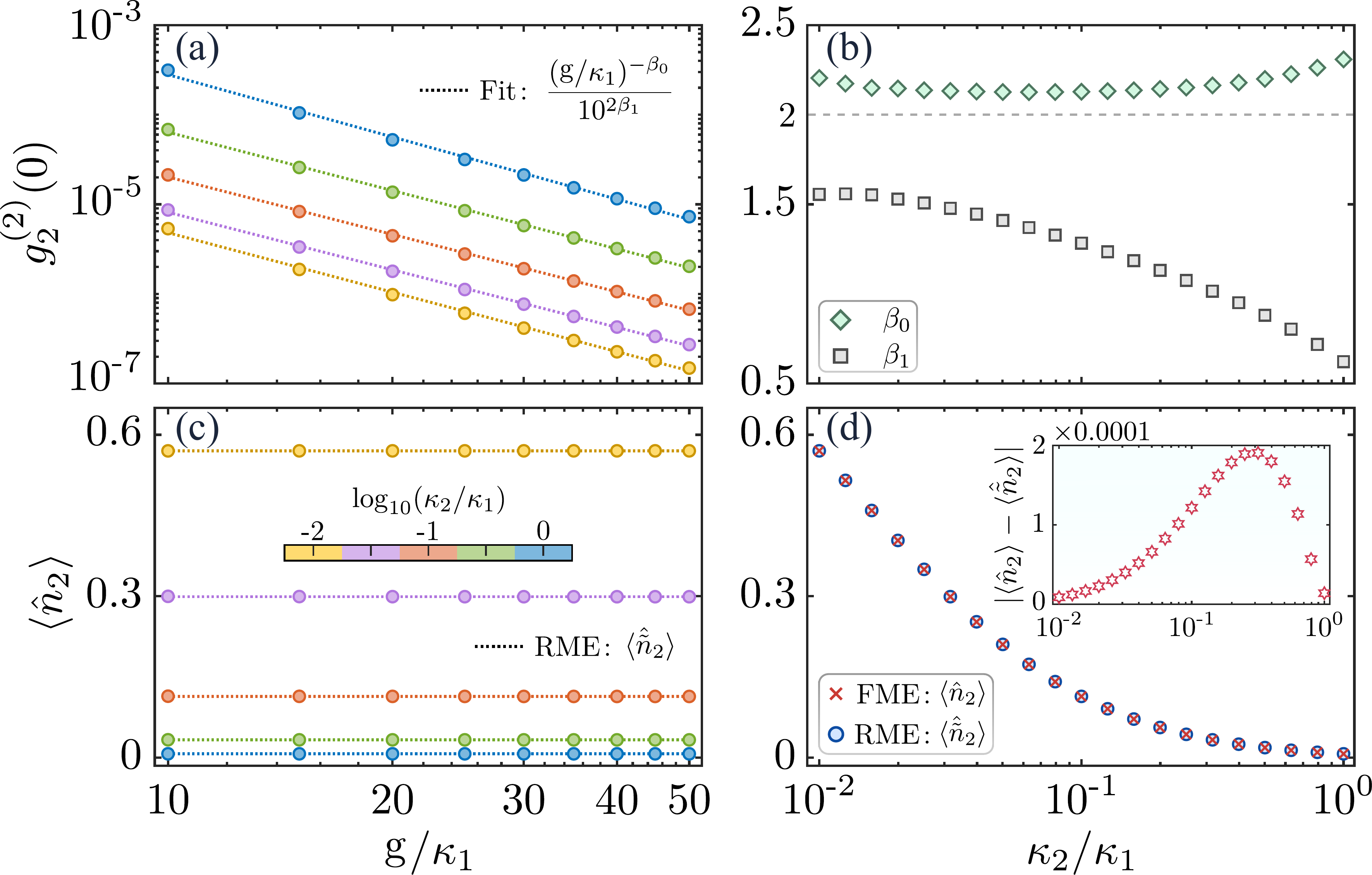}\\
        \caption{Panels (a) and (c) show the correlation $g^{(2)}_{2}(0)$ and the mean photon number $\expval{\hat{n}_2}$ versus the coupling strength $\newg/\kappa_1$ for different values of $\kappa_2/\kappa_1$ (encoded by color), respectively. Panel (b) presents the fitting coefficients $\beta_0$ and $\beta_1$, extracted from the power-law fit shown by the dashed lines in panel (a). (d) Validation of the RME against the FME at $\newg=20\kappa_1$ for computation of the mean photon number $\expval{\hat{n}_2}$ as a function of $\kappa_2/\kappa_1$, with the difference $|\!\expval{\hat{n}_2}-\langle\hat{\tilde{n}}_2\rangle|$ shown in the inset. Other parameters are $\Omega=0.5\kappa_1$, $J=0.1\kappa_1$, and $\gamma=0.01\kappa_1$.
        }\label{fig3}
    \end{figure}

    \emph{Beyond the weak driving limit}.---While the large suppression of $g_2^{(2)}(0)$ is interesting in the weak driving limit, the fact that the steady state corresponds to almost no photon in the cavity (i.e., $\expval{\hat{n}_2}\approx0$) makes the state inconvenient for applications \cite{senellart2017high, PhysRevA.90.063824, lpor.201900279}. To this end, it is necessary to go beyond the weak driving limit while strongly suppressing multiphoton populations, as this is a prerequisite for high-quality steady-state single-photon sources. Figure \ref{fig3}(a) shows that, beyond the weak driving regime (e.g., $\Omega/\kappa_1=0.5$), a near-perfect PB effect persists in the undriven cavity mode, and the second-order correlation also follows a clear power-law decay with increasing 3BI strength [see legend of Fig.\,\ref{fig3}(a)]. It is worth mentioning that this power-law behavior mainly originates from the $\order{J/\newg}$ correction term (see Appendix A in End Matter). Moreover, the fitted exponent $\beta_0$ is smaller than the biquadratic scaling observed in the weak driving limit, yet it remains above the quadratic scaling, i.e., $2<\beta_0<4$, as shown in Fig.\,\ref{fig3}(b). This thus indicates that the strong suppression of multiphoton populations in the undriven cavity mode persists even in the strong driving regime at sufficiently large 3BI strength. More importantly, we find that reducing the decay rate $\kappa_2$ not only further improves the single-photon purity, defined as $P=1-g_2^{(2)}(0)$, but also substantially increases the brightness $\expval{\hat{n}_2}$, as shown in Figs.\,\ref{fig3}(a) and \ref{fig3}(c). While achieving high purity and high brightness (i.e., $P\approx1$ and $\expval{\hat{n}_2}\ge0.5$) has remained elusive in theoretical studies of PB, our results demonstrate that such coexistence is indeed achievable.

\begin{figure}
        \includegraphics[width=8.5cm]{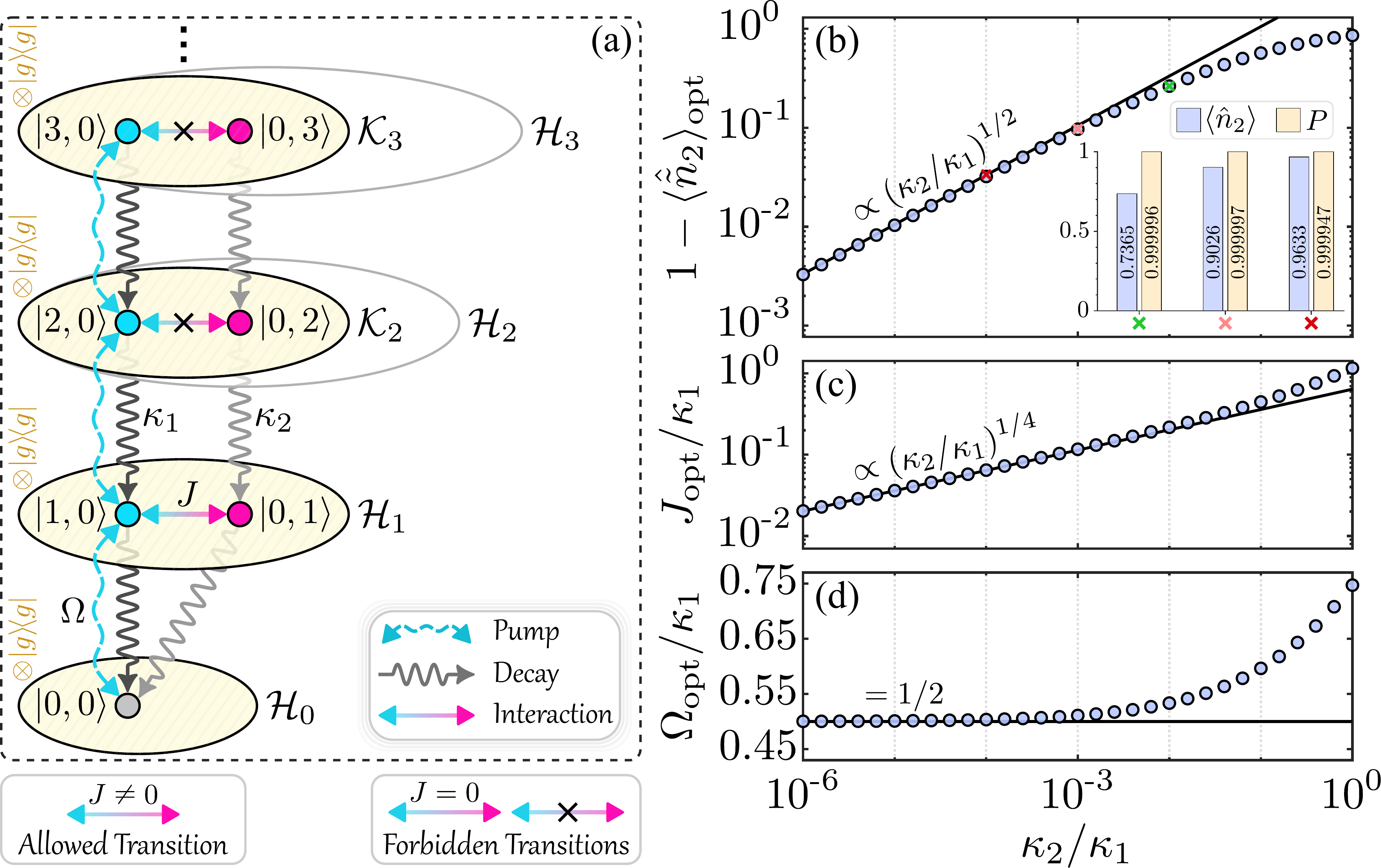}\\
        \caption{(a) Schematic illustration of the effective dynamics in the regime $J/\newg\ll1$ and $\aleph/\newg\ll1$ (or for $J=0$), including the driving, dissipation, and interaction processes. These colored dots represent the basis states that span the yellow regions. (b) Single-photon state infidelity $1-\langle\hat{\tilde{n}}_2\rangle_{\rm opt}$ versus the decay rate $\kappa_2/\kappa_1$ under the optimal parameters $J_{\rm opt}$ and $\Omega_{\rm opt}$ shown in panels (c) and (d). Notice that in panel (b), three discrete data points (crosses) are obtained from the FME at $\newg=20\kappa_1$ and $\gamma=0.01\kappa_1$ under the same optimal parameters, with the inset showing their values of $\expval{\hat{n}_2}$ and $P$. The solid black lines in panels (b)--(d) display the analytical asymptotic behaviors.
        }\label{fig4}
    \end{figure}

    \emph{High-brightness photon blockade and its mechanism}.---In Fig.\,\ref{fig3}, the undriven cavity mode exhibits near-perfect PB with markedly enhanced brightness. Figure \ref{fig4} further demonstrates that this PB persists even as the brightness approaches unity [see inset of Fig.\,\ref{fig4}(b)]. These results reveal a nontrivial and counterintuitive feature of our system, in stark contrast to previous schemes. In the following, we present a comprehensive qualitative and quantitative explanation of the distinctive blockade mechanism responsible for this behavior.

    First, in the weak driving limit, the system dynamics can be accurately captured within the low-excitation manifold and completely governed by an effective non-Hermitian (NH) Hamiltonian $\hat{H}_{\rm eff}=\hat{H}_{\rm tot}+\hat{H}_{\rm nh}$, where $\hat{H}_{\rm nh}=-(i/2)(\kappa_1\hat{n}_1+\kappa_2\hat{n}_2+\gamma\hat{\sigma}^\dagger\hat{\sigma})$ arises from the Lindblad dissipators. Beyond the weak driving regime, however, the multi-excitation manifold must be included and quantum jumps can no longer be neglected. Second, we explicitly confirm that in the absence of the 2BI ($J=0$), the interaction Hamiltonian (\ref{Eq1}) supports a doubly degenerate zero-energy subspace $\mathcal{K}_n$ that satisfies $\hat{H}_{\rm int}\ket{\psi}=0$ for all $\ket{\psi}\in{\mathcal{K}_n}$, where
    \begin{align}\label{Eq7}
        \mathcal{K}_n\coloneqq\mathrm{Span}\{|n,0\rangle|g\rangle,\ |0,n\rangle|g\rangle\}\subseteq\mathcal{H}_n\quad (n\ge1).
    \end{align}
    On the one hand, the basis states spanning the degenerate zero-energy subspace are also eigenstates of the NH term, satisfying $\hat{H}_{\rm nh}\ket{n,0}\ket{g}=-(i/2)n\kappa_1\ket{n,0}\ket{g}$ and $\hat{H}_{\rm nh}\ket{0,n}\ket{g}=-(i/2)n\kappa_2\ket{0,n}\ket{g}$. On the other hand, neither quantum jumps nor external drive induce leakage out of the zero-energy manifold $\mathcal{K}=\mathcal{H}_0\oplus\mathcal{K}_1\oplus\cdots\oplus\mathcal{K}_\infty$, as shown in Fig.\,\ref{fig4}(a); instead,
    they generate transitions only within the hierarchy $\{\mathcal{K}_n\}$, since $\{\hat{o}, \hat{H}_{\rm d}\}\ket{\psi}\in\mathcal{K}$ for all $\ket{\psi}\in\mathcal{K}$, where $\hat{o}\in\{\hat{a}_1,\hat{a}_2,\hat{\sigma}\}$ denotes the jump operator. Considering that $\mathcal{K}_1\subseteq\mathcal{H}_1$ and $\dim\mathcal{K}_1=\dim\mathcal{H}_1$, the zero-energy subspace $\mathcal{K}_1$ actually coincides with the single-excitation subspace $\mathcal{H}_1$, i.e., $\mathcal{K}_1=\mathcal{H}_1$. According to quantum trajectory theory \cite{Daley04032014}, if the system is initially prepared in a state $\ket{\psi}\in\mathcal{K}$, the ensuring dynamics remains confined to this manifold, i.e., $\hat{\rho}=\hat{\mathbb{P}}\hat{\rho}\mkern2mu\hat{\mathbb{P}}$, where $\mathbb{P}$ is the projection operator onto $\mathcal{K}$.

    Notably, in the presence of the 2BI ($J\neq 0$), we find that in the limit of $J/\newg\ll1$, the interaction Hamiltonian (\ref{Eq1}) still supports a pair of zero-energy eigenstates $\ket{0_\mathbf{k};2s}\simeq\ket{2s\delta_{\mathbf{k},1},2s\delta_{\mathbf{k},2}}\ket{g}$ in the even-excitation subspace $\mathcal{H}_{2s}$, as well as a pair of quasi-zero-energy eigenstates $\ket{\pm;2s+1}\simeq(\ket{2s+1,0}\pm\ket{0,2s+1})\ket{g}/\sqrt{2}$ in the odd-excitation subspace $\mathcal{H}_{2s+1}$ for $s\ge1$ [cf. Eq.~(\ref{Eq2})] (for the detailed derivation, see Supplemental Material \cite{supp}). This indicates that the subspaces $\mathcal{K}_n$ for all $n\ge2$ are approximately invariant under the interaction Hamiltonian. We notice that the coherent population transfer between the single-excitation states $\ket{1,0}\ket{g}$ and $\ket{0,1}\ket{g}$ is allowed due to the presence of the 2BI, as clearly seen in Fig.\,\ref{fig4}(a). To ensure that dissipation (the NH term) does not elevate the corrections of order $J/\newg$ neglected in the above asymptotic expressions to leading order, we further require $\aleph/\newg\ll1$, where $\aleph=\max\{\kappa_1,\kappa_2,\gamma\}$. As a result, we conclude that in the regime $J/\newg\ll1$ and $\aleph/\newg\ll1$, the system dynamics remains predominantly confined to the manifold $\mathcal{K}$, thereby resulting in $\hat{\rho}\simeq\hat{\mathbb{P}}\hat{\rho}\mkern2mu\hat{\mathbb{P}}$. The validity of this projection approximation is further verified numerically (see Appendix A in End Matter). For the sake of simplicity, we initially prepare the system in the state $\ket{G}\in\mathcal{H}_0$. Naturally, by projecting the FME (\ref{Eq4}) onto the manifold $\mathcal{K}$ and using the relation $\hat{\rho}\simeq\hat{\mathbb{P}}\hat{\rho}\mkern2mu\hat{\mathbb{P}}\equiv\hat{\tilde{\rho}}\otimes\ketbra{g}{g}$, we obtain an effective reduced master equation (RME)
    \begin{align}\label{Eq9}
        \dot{\hat{\tilde{\rho}}}=-i[\hat{\tilde{H}}_{\rm tot},\hat{\tilde{\rho}}]+\kappa_1\mathcal{D}[\hat{\tilde{a}}_1]\hat{\tilde{\rho}}+\kappa_2\mathcal{D}[\hat{\tilde{a}}_2]\hat{\tilde{\rho}}
    \end{align}
    with reduced Hamiltonian
    \begin{align}\label{Eq10}
        \hat{\tilde{H}}_{\rm tot}=J(\ketbra{1,0}{0,1}+\ketbra{0,1}{1,0})+\Omega(\hat{\tilde{a}}_1^{}+\hat{\tilde{a}}_1^\dagger),
    \end{align}
    where $\hat{\tilde{a}}_1=\hat{a}_1\otimes|0\rangle_2\langle0|$ and $\hat{\tilde{a}}_2=|0\rangle_1\langle0|\otimes\hat{a}_2$. In Eq.~(\ref{Eq10}), the first term indicates that the two cavity modes are coupled only in the single-excitation subspace $\mathcal{H}_1$, whereas they are effectively decoupled in the subspaces $\mathcal{K}_{n\neq1}$, as depicted by the arrows of allowed and forbidden transitions in Fig.\,\ref{fig4}(a). As a result, we obtain $\hat{\tilde{\rho}}|0,n\ge2\rangle=0$, thereby resulting in $\tilde{g}_2^{(2)}(0)=\langle\hat{\tilde{n}}_2(\hat{\tilde{n}}_2-1)\rangle/\langle\hat{\tilde{n}}_2\rangle^2=0$ and $\langle\hat{\tilde{n}}_2\rangle=\mel{0,1}{\hat{\tilde{\rho}}_{\rm ss}}{0,1}$, where $\hat{\tilde{n}}_2=\hat{\tilde{a}}_2^\dagger\hat{\tilde{a}}_2^{}$. It is worth noting that Fig.\,\ref{fig3}(d) and its inset present a comparison between $\langle\hat{\tilde{n}}_2\rangle$ and $\expval{\hat{n}_2}$, which exhibit excellent agreement between the two, thereby validating the RME.

    Having established the robust emergence of a perfect photon blockade in the undriven cavity mode within the RME framework, we now turn to enhancing the corresponding mean photon number $\langle\hat{\tilde{n}}_2\rangle$, i.e., the brightness. From Fig.\,\ref{fig4}(a), we clearly find that the target state $|0,1\rangle$ originates exclusively through coherent population transfer from $|1,0\rangle$, and incoherent loss processes only induce its decay into the vacuum state $|0,0\rangle$. As a result, reducing the decay rate $\kappa_2$ provides the most effective route to enhance the population of $\ket{0,1}$, as evidenced in Fig.\,\ref{fig4}(b), where the single-photon state infidelity decreases as $\kappa_2$ is reduced, exhibiting an asymptotic power-law scaling, i.e., $1-\langle\hat{\tilde{n}}_2\rangle_{\rm opt}\sim(\kappa_2/\kappa_1)^{1/2}$ (see Appendix A in End Matter). Figures \ref{fig4}(c) and \ref{fig4}(d) clearly show that for smaller $\kappa_2$, the optimal driving amplitude stabilizes at $\Omega_{\rm opt}=0.5\kappa_1$, and the optimal hopping strength tends to zero, following the scaling relation $J_{\rm opt}/\kappa_1\sim(\kappa_2/\kappa_1)^{1/4}$ (see End Matter). This behavior can be understood from the fact that maximizing the population of $\ket{1,0}$ at $J=0$ is approximately equivalent to optimizing the population of $\ket{0,1}$ at weak hopping $J/\kappa_1\ll1$, owing to coherent population transfer and perturbation theory. Specifically, the steady state of the system at $J=0$ can be obtained by analytically solving the RME, i.e., $\hat{\tilde{\rho}}_{\rm ss}=|\alpha\rangle_1\langle\alpha|\otimes|0\rangle_2\langle0|$, where $|\alpha\rangle_1$ is the coherent state with amplitude $\alpha=-2i\Omega/\kappa_1$. Then, by maximizing $\mel{1,0}{\hat{\tilde{\rho}}_{\rm ss}}{1,0}$, one obtains the optimal condition $\abs{\alpha}=1$, i.e., $\abs{\Omega_{\rm opt}}=0.5\kappa_1$. 

    Finally, we would like to comment on the feasibility of the system parameters and the experimental implementation. Specifically, Fig.\,\ref{fig3}(a) and Fig.\,\ref{fig4}(b) clearly show that near-perfect purity relies on a strong 3BI strength, and near-unity brightness requires a very weak decay rate $\kappa_2$. In fact, the former can be substantially enhanced via several feasible approaches, such as squeezing-induced exponential enhancement \cite{PhysRevLett.130.073602, pan2025, cbrb-8xkh} and multi-atom collective enhancement \cite{RevModPhys.82.1041, Guerin15052017, singh2013dynamics}, whereas the latter can be facilitated by introducing a two-photon dissipation, thereby greatly relaxing the stringent requirement on $\kappa_2$; see Appendix B in End Matter. Moreover, we present a feasible superconducting-circuit implementation of our scheme in the Supplemental Material \cite{supp}.
    
    \emph{Conclusions}.---We have proposed an extended nondegenerate two-photon Jaynes-Cummings model featuring coexisting two-body and three-body interactions. In this system, we have identified a novel mechanism that allows near-perfect photon blockade of bosonic modes to be realized without sacrificing brightness, i.e., with a near-unity mean photon number. This mechanism is underpinned by the distinctive energy-level structure arising from the combined action of the two-body and three-body interactions. In particular, we have also theoretically derived an effective reduced master equation, which provides a clear and intuitive physical picture for the coexistence of near-perfect photon blockade and high brightness. Our work not only overcomes this outstanding challenge in theoretical studies of photon blockade but also opens a promising avenue toward near-ideal single-photon sources with potential applications in quantum information, communication, and metrology.

    \emph{Acknowledgments}.---This work is supported by the National Science Fund for Distinguished Young Scholars of China (Grant No. 12425502), the National Key Research and Development Program of China (Grant No. 2021YFA1400700), and the Fundamental Research Funds for the Central Universities (Grant No. 2024BRA001).

    \emph{Data availability.}---The data that support the findings of this Letter are not publicly available. The data are available from the authors upon reasonable request.

    \bibliography{reference}

    
    \onecolumngrid
    \vspace{1em}
    \begin{center}
        \textbf{\Large End Matter}
    \end{center}
    \vspace{1em}
    \twocolumngrid

    \begin{figure}[t]
        \includegraphics[width=8.5cm]{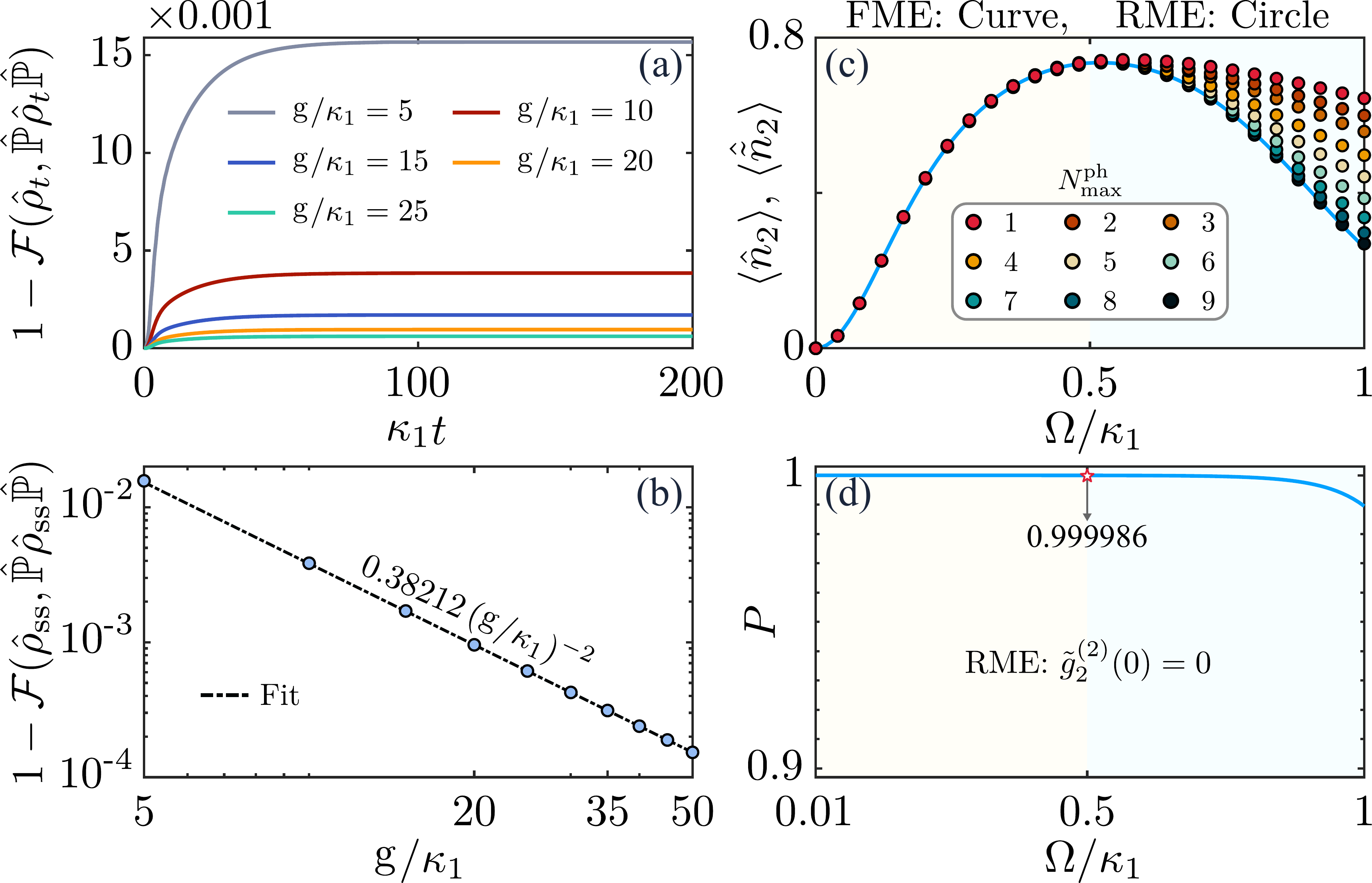}\\
        \caption{Panel (a) shows the infidelity dynamics of the projection approximation for different 3BI strengths, while panel (b) displays its steady-state infidelity dependence on the 3BI strength. Panels (c) and (d) show the mean photon numbers $\expval{\hat{n}_2}$ and $\langle\hat{\tilde{n}}_2\rangle$ (for $N_{\rm max}^{\rm ph}=1\sim9$), and single-photon purity $P$ as functions of the driving amplitude $\Omega/\kappa_1$, respectively. Note that we have used $\Omega=0.5\kappa_1$ in panels (a)--(b) and $\newg=10\kappa_1$ in panels (c)--(d). The remaining parameters are $\gamma=0.01\kappa_1$, $\kappa_2=0.01\kappa_1$, and $J=J_{\rm opt}$ [see Eq.~(\ref{Eq18})].
        }\label{fig5}
    \end{figure}

    \emph{Appendix A: Validation and analytical solution of the reduced master equation}.---In this section, we systematically establish the validity of the RME and derive the asymptotic scaling observed in Figs.\,\ref{fig4}(b) and \ref{fig4}(c). First, we note that the RME is derived from the FME by employing the projection approximation $\hat{\rho}\simeq\hat{\mathbb{P}}\hat{\rho}\mkern2mu\hat{\mathbb{P}}$. This implies that the better the approximation holds, the smaller the difference between the RME and FME results. To quantitatively assess the accuracy of this approximation, we introduce the fidelity between the full density matrix and its projected counterpart, which is defined as 
    \begin{align}\label{Eq11}
        \mathcal{F}(\hat{\rho}, \hat{\mathbb{P}}\hat{\rho}\mkern2mu\hat{\mathbb{P}})=\left[\Tr(\sqrt{\sqrt{\hat{\rho}}\mkern2mu\hat{\mathbb{P}}\hat{\rho}\mkern2mu\hat{\mathbb{P}}\sqrt{\hat{\rho}}})\right]^2.
    \end{align}
   In Fig.\,\ref{fig5}(a), we find that the infidelity increases monotonically with time and saturates at a steady-state value, which remains well below unity. Besides, the steady-state infidelity is substantially suppressed as the 3BI strength increases and exhibits a clear power-law scaling behavior, i.e., $1-\mathcal{F}(\hat{\rho}_{\rm ss}, \hat{\mathbb{P}}\hat{\rho}_{\rm ss}\mkern2mu\hat{\mathbb{P}})\sim(J/\newg)^{2}$, as depicted in Fig.\,\ref{fig5}(b). As a result, by defining $\|\hat{\mathcal{O}}(J/\newg)\|_1=\order{J/\newg}$, we have $\hat{\rho}=\hat{\mathbb{P}}\hat{\rho}\mkern2mu\hat{\mathbb{P}}+\hat{\mathcal{O}}(J/\newg)$, following from the inequality \cite{761271}
   \begin{align}\label{Eq12}
       \|\hat{\rho}-\hat{\mathbb{P}}\hat{\rho}\mkern2mu\hat{\mathbb{P}}\|_1\le2\big[1-\mathcal{F}(\hat{\rho}, \hat{\mathbb{P}}\hat{\rho}\mkern2mu\hat{\mathbb{P}})\big]^{1/2},
   \end{align}
    where $\norm{\bigcdot}_1$ denotes the trace norm. This correction term $\hat{\mathcal{O}}(J/\newg)$ not only directly explains the power-law behavior observed in Fig.\,\ref{fig3}(a), but also quantitatively confirms the accuracy of the projection approximation. Consequently, in the regime $J/\newg\ll1$ and $\aleph/\newg\ll1$, the RME provides a more intuitive and accurate description of the system dynamics. Second, we also note that the density matrix in the MRE satisfies $\hat{\tilde{\rho}}\ket{0,n\ge2}=0$. This indicates that the undriven cavity mode does not support multiphoton occupation, so that its Hilbert space can be consistently truncated to Fock states with at most one photon. For the driven cavity mode, its photonic Hilbert space is not intrinsically constrained and is therefore truncated to Fock states with at most $N_{\rm max}^{\rm ph}$ photons for numerical and analytical convenience. As a result, the RME effectively restricts the system dynamics to a Hilbert space of dimension $N_{\rm max}^{\rm ph}+2$, spanned by the basis states
    \begin{align}\label{Eq13}
        \{|0,0\rangle, |0,1\rangle, |1,0\rangle, |2,0\rangle, \ldots,|N_{\rm max}^{\rm ph},0\rangle\}\subset\mathcal{K}.
    \end{align}
    It should be noted that in our simulation, all numerical results obtained from the FME are computed by truncating the photonic Hilbert spaces of the two cavity modes to Fock states with at most ten photons, whereas for all results obtained from the RME we set $N_{\rm max}^{\rm ph}=30$, unless otherwise specified.

    Figure \ref{fig5}(c) presents a comparison between $\langle\hat{\tilde{n}}_2\rangle$ for different $N_{\rm max}^{\rm ph}$ and $\expval{\hat{n}_2}$ as functions of the driving amplitude. Interestingly, we find that in the regime $\Omega\le0.5\kappa_1$, the results obtained from the RME at $N_{\rm max}^{\rm ph}=1$ agree perfectly with those from the FME, and the driving amplitude corresponding to the maximal mean photon number also lies within this regime. By contrast, as the driving amplitude is further increased (i.e., $\Omega>0.5\kappa_1$), not only is a larger value of $N_{\rm max}^{\rm ph}$ correspondingly required, but also both the mean photon number and the single-photon purity gradually decrease [see Figs.\,\ref{fig5}(c) and \ref{fig5}(d)]. Therefore, since the optimal driving amplitude stabilizes at $\Omega_{\rm opt}=0.5\kappa_1$ [see Fig.\,\ref{fig4}(d)], setting $N_{\rm max}^{\rm ph}=1$ is sufficient and well justified for analyzing the asymptotic behaviors observed in Figs.\,\ref{fig4}(b) and \ref{fig4}(c). In this setting, the RME (\ref{Eq9}) restricts the system dynamics to an analytically solvable three-level system, taking the form 
    \begin{align}\label{Eq14}
        \dot{\hat{\tilde{\rho}}}=-i[\hat{\tilde{H}}_{\rm tot},\hat{\tilde{\rho}}]+\kappa_1\mathcal{D}[\ketbra{0}{L}]\hat{\tilde{\rho}}+\kappa_2\mathcal{D}[\ketbra{0}{R}]\hat{\tilde{\rho}},
    \end{align}
    where
    \begin{align}\label{Eq15}
        \hat{\tilde{H}}_{\rm tot}=J(\ketbra{L}{R}+\ketbra{R}{L})+\Omega(\ketbra{0}{L}+\ketbra{L}{0}),
    \end{align}
    Here we denote $\ket{0}\equiv\ket{0,0}$, $\ket{L}\equiv\ket{1,0}$, and $\ket{R}\equiv\ket{0,1}$ as the three-levels of the effective system. Then, by analytically solving Eq.~(\ref{Eq14}) in the basis states $\{|0\rangle,|L\rangle,|R\rangle\}$, we obtain the steady-state density matrix $\tilde{\bm\rho}_{\rm ss}$, which can be represented in matrix form as 
    \begin{align}\label{Eq16}
        \tilde{\bm\rho}_{\rm ss}=\begin{pmatrix}
            \tilde{\rho}_{00}^{} & \tilde{\rho}_{0L}^{} & \tilde{\rho}_{0R}^{}\\
            \tilde{\rho}_{L0}^{} & \tilde{\rho}_{LL}^{} & \tilde{\rho}_{LR}^{}\\
            \tilde{\rho}_{R0}^{} & \tilde{\rho}_{RL}^{} & \tilde{\rho}_{RR}^{}
        \end{pmatrix}.
    \end{align}
    Thus, the mean photon numbers of the driven and undriven cavity modes can be obtained as \cite{supp}
    \begin{subequations}
    \begin{align}
        \langle\hat{\tilde{n}}_1\rangle&=\tilde{\rho}_{LL}^{}=\frac{4\mathcal{Q}_2\Omega^2(\kappa_1^2+8\Omega^2)^{-1}}{16\mathcal{Q}_0J^4+8\mathcal{Q}_1J^2+\mathcal{Q}_2},\label{Eq17a}\\
        \langle\hat{\tilde{n}}_2\rangle&=\tilde{\rho}_{RR}^{}=\frac{16\mathcal{Q}_0J^2\Omega^2}{16\mathcal{Q}_0J^4+8\mathcal{Q}_1J^2+\mathcal{Q}_2},\label{Eq17b}
    \end{align}
    \end{subequations}
    where $\mathcal{Q}_0=\kappa_1+\kappa_2$, $\mathcal{Q}_1=\kappa_1(2\Omega^2+\kappa_1\kappa_2+\kappa_2\kappa_2)$, and $\mathcal{Q}_2=\kappa_2(\kappa_1^2+8\Omega^2)(4\Omega^2+\kappa_1\kappa_2+\kappa_2\kappa_2)$. For clarity, we denote the single-photon state infidelity as $\mathcal{I}=1-\langle\hat{\tilde{n}}_2\rangle$. Next, we substitute the driving amplitude $\Omega$ in Eq.~(\ref{Eq17b}) with its optimal value $\Omega_{\rm opt}=0.5\kappa_1$. Finally, the optimal hopping amplitude is determined by solving $\partial\mathcal{I}/\partial J=0$, which results in 
    \begin{align}\label{Eq18}
        J_{\rm opt}=\kappa_1\left[\frac{3\kappa_2}{16\kappa_1}\left(\frac{\kappa_1+\kappa_2}{\kappa_1}-\frac{\kappa_2}{\kappa_1+\kappa_2}\right)\right]^{1/4}.
    \end{align}
    Accordingly, the optimal infidelity is computed as
    \begin{align}\label{Eq19}
        \mathcal{I}_{\rm opt}=\frac{\epsilon(1+2\epsilon)+\sqrt{12\epsilon(1+\epsilon)(1+\epsilon+\epsilon^2)}}{1+2\epsilon(1+\epsilon)+\sqrt{12\epsilon(1+\epsilon)(1+\epsilon+\epsilon^2)}},
    \end{align}
    where $\epsilon=\kappa_2/\kappa_1$. To obtain the asymptotic scaling, we expand Eqs.~(\ref{Eq18})--(\ref{Eq19}) in powers of $\epsilon$ as $\epsilon\to0$, yielding
    \begin{subequations}
    \begin{align}
        J_{\rm opt}/\kappa_1&=(3\epsilon/16)^{1/4}+\mathcal{O}(\epsilon^{9/4})\sim(\kappa_2/\kappa_1)^{1/4},\label{Eq20a}\\
        \mathcal{I}_{\rm opt}&=(12\epsilon)^{1/2}+\order{\epsilon}\sim(\kappa_2/\kappa_1)^{1/2}.\label{Eq20b}
    \end{align}
    \end{subequations}

    \begin{figure}[t]
        \includegraphics[width=8.5cm]{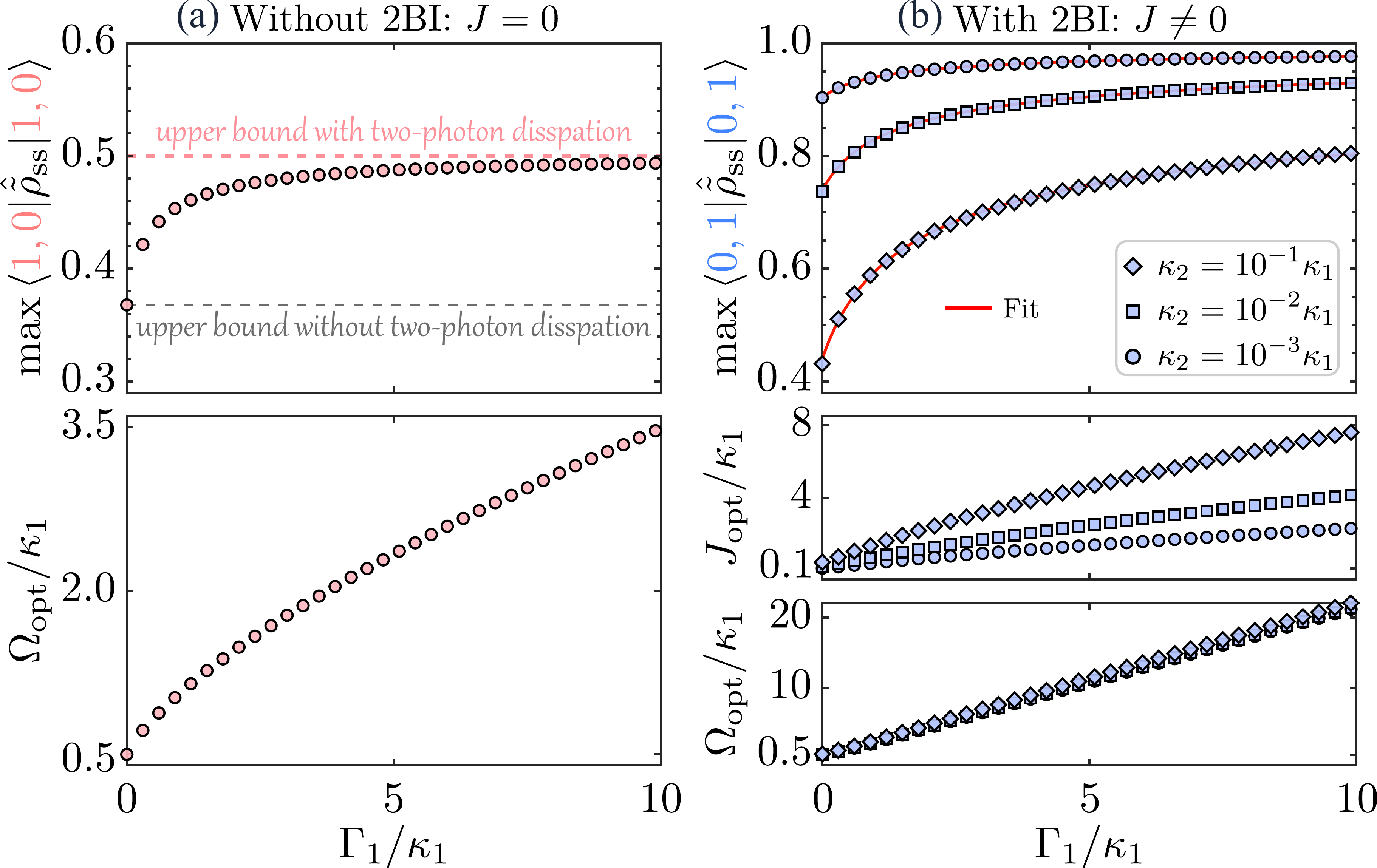}\\
        \caption{Maximal single-photon populations (a) $\mel{1,0}{\hat{\tilde{\rho}}_{\rm ss}}{1,0}$ without 2BI ($J=0$) and (b) $\mel{0,1}{\hat{\tilde{\rho}}_{\rm ss}}{0,1}$ with 2BI ($J\neq0$) versus the two-photon decay rate $\Gamma_1/\kappa_1$ for different values of $\kappa_2=\{0.1,0.01,0.001\}\kappa_1$. The corresponding optimal parameters are shown in the bottom panels. The solid red curves in panel (b) are fitted to the form $1-p(1+q\Gamma_1/\kappa_1)^{-\nu}$, where the fitting parameters $p$, $q$, and $\nu$ are summarized in Table~\ref{table1}.
        }\label{fig6}
    \end{figure}
    
    \emph{Appendix B: Role of two-photon dissipation}.---While we demonstrate that the coexistence of near-perfect photon blockade and a near-unity mean photon number is achievable, attaining such performance requires a low decay rate $\kappa_2$ [cf. Eq.~(\ref{Eq20b})], which typically necessitates a high cavity quality factor. To alleviate this experimental constraint, a simple and feasible method is to introduce two-photon loss channels into our system, which is equivalent to adding the following terms into the FME (\ref{Eq4}):
    \begin{align}\label{Eq21}
        \Gamma_1\mathcal{D}[\hat{a}_1^2]\hat{\rho}+\Gamma_2\mathcal{D}[\hat{a}_2^2]\hat{\rho},
    \end{align}
    where $\Gamma_{1}$ and $\Gamma_2$ are the two-photon decay rates of the driven and undriven cavity modes, respectively. Notably, the introduction of two-photon loss channels does not invalidate the projection approximation established in the main text. Accordingly, the RME including two-photon dissipation can be obtained by adding the two terms in Eq.~(\ref{Eq21}) to Eq.~(\ref{Eq9}) and replacing $\hat{a}_{\bf k}$ by $\hat{\tilde{a}}_{\bf k}$. Remarkably, two-photon dissipation restores the biquadratic scaling of $g_2^{(2)}(0)$ with increasing 3BI strength (for the detailed discussion, see Supplemental Material \cite{supp}). By contrast, the intrinsic structure of the RME implies that only the two-photon dissipation of the driven cavity mode contributes to $\expval{\hat{n}_2}$. Hence, in the following, we consider only the two-photon dissipation of the driven cavity mode.

    \begin{table}[b]
    \caption{Fitting parameters $p$, $q$, and $\nu$ as functions of $\epsilon$.}
    \label{table1}
    \begin{ruledtabular}
    \begin{tabular}{cccc}
    $\epsilon\ (\downarrow)$ & $p\ (\downarrow)$ & $q\ (\uparrow)$ & $\nu\ (\uparrow)$\\
    \hline
    \rule{0pt}{2.5ex}0.1   & 0.5604 & 1.2521 & 0.4038 \\
    0.01  & 0.2615 & 1.4772 & 0.4770 \\
    0.001 & 0.0964 & 1.5557 & 0.5089
    \end{tabular}
    \end{ruledtabular}
    \end{table}

    In Fig.\,\ref{fig6}(b), we find that $\max\mel{0,1}{\hat{\tilde{\rho}}_{\rm ss}}{0,1}$ increases monotonically with the two-photon dissipation and can be well described by the form $1-p(1+q\Gamma_1/\kappa_1)^{-\nu}$, while the optimal parameters $J_{\rm opt}$ and $\Omega_{\rm opt}$ show an approximately linear dependence on the two-photon dissipation. Moreover, as the single-photon dissipation $\kappa_2$ decreases, $J_{\rm opt}$ is overall reduced, and $\Omega_{\rm opt}$ gradually saturates, consistent with the behavior observed in Figs.\,\ref{fig4}(c) and \ref{fig4}(d). Physically, the two-photon-induced enhancement originates from the fact that two-photon dissipation increases the population of $\ket{1,0}$ at $J=0$. Specifically, in the absence of the 2BI, one has $\hat{\tilde{\rho}}_{\rm ss}=|\alpha\rangle_1\langle\alpha|\otimes|0\rangle_2\langle0|$, and the upper bound is given by $\mel{1,0}{\hat{\tilde{\rho}}_{\rm ss}}{1,0}=1/e$ at ${\abs{\alpha}=1}$. However, with the introduction of two-photon dissipation, multiphoton occupations of the driven cavity mode are suppressed, thereby resulting in an enhancement of this upper bound, as depicted in Fig.\,\ref{fig6}(a). In particular, in the limit $\Gamma_1\gg\kappa_1$, the system dynamics is effectively confined to the three-level system, such that the upper bound reaches $0.5$ upon maximizing Eq.~(\ref{Eq17a}) at $J=0$. As a result, the enhancement of the upper bound provides an indirect explanation for why introducing two-photon dissipation can further increase the population of the target state $\ket{0,1}$. These results demonstrate that introducing two-photon dissipation can substantially alleviate the stringent requirement on single-photon dissipation, thereby enhancing the experimental feasibility.

\end{document}